\documentstyle[aps,prl,twocolumn,epsfig,pstcol]{revtex}
\newcommand{\etal}{{\em et al.}}                %       et al.

\newcommand{\dn}[2]{d^{#1}{#2}\,}
\newcommand{\eqref}[1]{(\ref{#1})}

\begin{document}
\voffset0.5cm
\title{Model-independent source imaging using two-pion correlations\\
in 2 to 8$A$ GeV Au + Au collisions}
\author{S.Y.~Panitkin,$^7$
N.N.~Ajitanand,$^{12}$ J.~Alexander,$^{12}$ M.~Anderson,$^5$
D.~Best,$^1$ F.P.~Brady,$^5$ T.~Case,$^1$ W.~Caskey,$^5$
D.~Cebra,$^5$ J.~Chance,$^5$ P.~Chung,$^{12}$
B.~Cole,$^4$ K.~Crowe,$^1$ A.~Das,$^{10}$
J.~Draper,$^5$ M.~Gilkes,$^{11}$ S.~Gushue,$^2$
M.~Heffner,$^5$ A.~Hirsch,$^{11}$
E.~Hjort,$^{11}$ L.~Huo,$^6$ M.~Justice,$^7$
M.~Kaplan,$^3$ D.~Keane,$^7$ J. Kintner,$^8$
J.~Klay,$^5$ D.~Krofcheck,$^9$ R.~Lacey,$^{12}$ J.~Lauret,$^{12}$
M.A.~Lisa,$^{10}$
H.~Liu,$^7$ Y.M.~Liu,$^6$ R.~McGrath,$^{12}$ Z.~Milosevich,$^3$,
G.~Odyniec,$^1$ D.~Olson,$^1$  C.~Pinkenburg,$^{12}$
N.~Porile,$^{11}$ G.~Rai,$^1$ H.G.~Ritter,$^1$
J.~Romero,$^5$ R.~Scharenberg,$^{11}$ L.S.~Schroeder,$^1$
B.~Srivastava,$^{11}$ N.T.B.~Stone,$^2$ T.J.M.~Symons,$^1$
S.~Wang,$^7$ R.~Wells,$^{10}$ J.~Whitfield,$^3$ T.~Wienold,$^1$ R.~Witt,$^7$
L.~Wood,$^5$ X.~Yang,$^4$ W.N.~Zhang,$^6$ Y.~Zhang $^4$\\
(E895 Collaboration) \\
and D.A.~Brown$^{13}$, P. Danielewicz$^{14}$
}
\address{
$^1$Lawrence Berkeley National Laboratory, Berkeley, California 94720 \\
$^2$Brookhaven National Laboratory, Upton, New York 11973 \\
$^3$Carnegie Mellon University, Pittsburgh, Pennsylvania 15213 \\
$^4$Columbia University, New York, New York 10027 \\
$^5$University of California, Davis, California 95616 \\
$^6$Harbin Institute of Technology, Harbin 150001, P.~R.~China \\
$^7$Kent State University, Kent, Ohio 44242 \\
$^8$St.~Mary's College of California, Moraga, California 94575 \\
$^9$University of Auckland, Auckland, New Zealand \\
$^{10}$The Ohio State University, Columbus, Ohio 43210 \\
$^{11}$Purdue University, West Lafayette, Indiana 47907 \\
$^{12}$State University of New York, Stony Brook, New York 11794 \\
$^{13}$University of Washington, Seattle, Washington 98195 \\
$^{14}$Michigan State University, East Lansing, Michigan 48824\\
}
\date{\today}
\maketitle
\begin{abstract}
We report a particle source imaging analysis based on two-pion
correlations in high multiplicity Au + Au collisions at beam
energies between 2 and 8$A$ GeV.  We apply the imaging technique
introduced by Brown and Danielewicz, which allows 
a model-independent extraction of 
source functions with useful accuracy out to relative
pion separations of about 20 fm. The extracted source functions have
Gaussian shapes.
Values of source functions at zero separation are almost constant
across the energy range under study.
Imaging results are found to be 
consistent with conventional source parameters obtained from a 
multidimensional HBT analysis.
\end{abstract}
%PACS numbers here

One of the robust predictions of Quantum Chromodynamics is that strongly
interacting matter can exist in a state with colored degrees of freedom,
i.e., quarks and gluons, if subjected to sufficiently high temperature or
density~\cite{shuryak_80}.  This deconfined state of matter, often called
Quark-Gluon Plasma (QGP), may have been present in the Universe a few
microseconds after the Big Bang.  It is also believed that QGP can be
produced in collisions of heavy ions at high energies.  Experimental
searches for evidence of QGP are prominent and challenging priorities
in modern heavy-ion physics.  A major difficulty is that the sought-after
QGP is a transient state which persists only for timescales on the order
of fm$/c$.  After expansion and cooling, any QGP necessarily undergoes
a phase transition to hadronic matter, possibly leaving little
evidence behind in the final state of the collision.  Numerous
experimental observables have been proposed as signatures of QGP
creation in heavy-ion collisions~\cite{muller_99}.  One of the more
promising of these is based on the expectation that the larger
number of degrees of freedom associated with the deconfined state
should manifest itself in increased system entropy, which ought to
survive subsequent hadronization and freeze-out.  The specific
prediction is that QGP formation will cause the system to expand more
and/or interact for a longer time and produce more particles, relative
to a system which remained in the hadronic phase throughout the collision
process.

Intensity interferometry~\cite{gkp_71} has been extensively used to
extract information about spatial-temporal properties of heavy-ion
collisions.  The HBT
approach~\cite{gkp_71,pratt_90,gelbke_90,pratt_98,wiedemann_99,heinz_99}
is well understood from the theoretical point of view, and is the
mainstream analysis method for meson ($\pi, K$) intensity interferometry.
Meson final-state interactions are simple, so one may remove most of
the distortion of the final-state wavefunctions with a simple Coulomb
correction.  Furthermore, in the
traditional analysis one does not ask for more than the widths of the
underlying source function.
A novel approach --- the imaging technique introduced by Brown and
Danielewicz~\cite{dbrown_1,dbrown_2,panitkin_99_1} --- offers the 
opportunity to use {\it any} class of like-pair correlations to 
reconstruct the entire source function, as opposed to the conventional 
model-dependent parametrization.  
The two approaches have natural connections, which are 
especially clear for the case of non-interacting spin-zero 
bosons~\cite{dbrown_00_1}.  In this paper, we report the first
application of the source imaging technique to experimental data,
specifically negative pions from Au + Au collisions at beam energies
between 2 and 8$A$ GeV, with particular emphasis on verification and
study of the expected connection with the source parameters obtained
from HBT analysis.  Comparing the two methods not only allows
us to test the model underlying the HBT parametrization, but also lends
credibility to the source imaging approach in other contexts such as
proton-proton correlations.

We present data from experiment E895~\cite{rai_93}, in which Au beams
of kinetic energy 1.85, 3.9, 5.9 and 7.9$A$ GeV from the AGS accelerator
at Brookhaven were incident on a fixed Au target.  The analyzed
$\pi^-$ samples come from the main E895 subsystem --- the EOS Time
Projection Chamber (TPC)~\cite{rai_90}, located in a dipole magnetic
field of 0.75 or 1.0 Tesla.   Results of the E895
$\pi^-$ HBT analysis were reported earlier~\cite{lisa_1}, and in this
Letter, we focus on an application of imaging to the same datasets.

In order to obtain the two-pion correlation function $C_2$
experimentally, the standard event-mixing technique was 
used~\cite{kopylov_74}. 
 Negative pions were detected and
reconstructed over a substantial fraction of $4\pi$ solid angle in
the center-of-mass frame, and simultaneous measurement of particle
momentum and specific ionization in the TPC gas helped to separate
$\pi^-$ from other negatively charge particles, such as electrons,
K$^-$ and antiprotons.  Contamination from these species is estimated
to be under 5\%, and is even less at the lower beam energies and
higher transverse momenta.  Momentum resolution in the region of 
correlation measurements is better than 3\%.
Event centrality selection was based on the multiplicity of 
reconstructed charged particles.  In the present analysis, events were
selected with a multiplicity corresponding to the upper 11\% of
the inelastic cross section for Au + Au collisions.  Only pions with 
$p_T$ between 100 MeV/$c$ and 300 MeV/$c$ and within $\pm 0.35$ units 
from midrapidity were used.
Another requirement was for each used $\pi^-$ track to point back to the primary 
event vertex with a distance of closest approach (DCA) less than 2.5 cm.
This cut removes most pions originating from weak decays of long-lived
particles, e.g. $\Lambda$ and K$^0$.  Monte Carlo simulations based on 
the RQMD model~\cite{sorge_95} indicate that decay daughters are present 
at a level that varies from 5\% at 4$A$ GeV to 10\% at 8$A$ GeV, and   
they lie preferentially at $p_T < 100$ MeV/$c$.  
Finally, in order to overcome effects of track merging, a cut on spatial
separation of two tracks was imposed.  For pairs from both ``true'' and 
``background'' distributions, the separation between two tracks was 
required to be greater than 4.5 cm over a distance of 18 cm in the beam 
direction. This cut also suppresses effects of track
splitting~\cite{lisa_1}.

Besides such cuts, the correlation functions themselves were
corrected for Coulomb final-state interactions and finite momentum
resolution effects, using an iterative technique similar to that
used by the NA44 Collaboration [19].  See [16] for details.

Imaging~\cite{dbrown_1,dbrown_2} was used to extract the shape of 
the source from the measured correlation functions. It has been
shown~\cite{panitkin_99_1} that  
imaging allows robust reconstruction of the source function for systems 
with non-zero lifetime, even in the presence of strong space-momentum
correlations.  The main features of the method are outlined below; see 
Refs~\cite{dbrown_1,dbrown_2,dbrown_00_1} for more details.  The 
full two-particle correlation function may be expressed as: 
\begin{equation}
        C_{\bf P}({\bf Q}) -1 =
        \int d{\bf r} \,K({\bf Q}, {\bf r}) \, S_{\bf P} ({\bf r}) \,,
        \label{panitkin_K}
\end{equation}
where $K = |\Phi_{\bf Q}^{(-)}({\bf r})|^2-1$, the total momentum of the 
pair is ${\bf P}$, the relative momentum is ${\bf Q}={\bf p_1}-{\bf p_2}$, 
and the relative separation of emission points 
is ${\bf r}$.  Both ${\bf r}$ and ${\bf Q}$ are written in the pair center of
mass frame.  Here, $\Phi_{\bf Q}^{(-)}$ is the 
relative wavefunction of the pair and for spin-$0$ bosons in the 
absence of final-state interactions, it is the symetrized sum of free waves:
\begin{equation}
        \Phi_{{\bf Q}}^{(-)}({\bf r})=\frac{1}{\sqrt{2}}\left
         ( e^{i{\bf Q}/2\cdot{\bf r}} +e^{-i {\bf Q}/2\cdot {\bf r}}\right).
\end{equation}
However, in general it may be quite complicated.
The goal of imaging is to the determine the relative
source function ($S_{\bf P} ({\bf r})$ in Eq.~(\ref{panitkin_K})), 
given $C_{\bf P}({\bf Q})$.  
The problem of imaging then becomes the
problem of inverting $K({\bf q}, {\bf r})$.
The source function, $S_{\bf P} ({\bf r})$,
is the distribution of relative separations of emission points for the two 
particles in their center-of-mass frame. 
  
In this letter, we analyze the pion correlations in $Q_{inv}$.  The 
transition from the full three-dimensional problem in Eq.~\eqref{panitkin_K} 
to the angle-averaged problem is straightforward: one expands the source and
correlation in spherical harmonics~\cite{dbrown_1} and keep $\ell=m=0$
terms.
The angle-averaged version of Eq.~\eqref{panitkin_K} is
\begin{equation}
   C(Q_{inv})-1=4\pi \int \dn{}{r}r^2 K(Q_{inv},r) S(r).
\end{equation}
Here $Q_{inv}=\sqrt{{\bf Q}^2-Q_0^2}$ and since $Q_0=0$ in the pair 
CM frame, $Q_{inv}=|{\bf Q}|$.  
Thus, while the source one reconstructs is a distribution in the pair
CM frame, the inversion itself may be done with correlations
constructed in any frame. 
In this equation, the kernel is simply averaged over the angle between 
${\bf q}$ and ${\bf r}$:
\begin{equation}
   K(Q_{inv},r)=\frac{1}{2}\int^1_{-1}\dn{}{(\cos\theta_{\bf Qr})} 
   K({\bf Q},{\bf r}).
\end{equation}
For identical spin-zero bosons with no FSI, this kernel is 
\begin{equation}
   K(Q_{inv},r)=\sin{(Q_{inv}r)}/Q_{inv}r.
   \label{eqn:noFSIkernel}
\end{equation}
We comment that since the ${\bf r}=0$ point of the source is a maximum, it is
spherically symmetric in a neighborhood around that point and 
$S(r\rightarrow 0)=S({\bf r}\rightarrow 0)$.
We now proceed as in \cite{dbrown_in_prep} and 
expand the radial dependence of the source in a Basis Spline 
basis: $S({\bf r})=\sum_j S_j B_j({\bf r})$.
With this, Eq.~\eqref{panitkin_K} becomes a matrix equation
$C_i=\sum_j K_{ij} S_j$ with a new kernel:
\begin{equation}
        K_{ij}=\int d{\bf r} K({\bf Q}_i,{\bf r}) B_j({\bf r}).  
\end{equation}
Imaging reduces to finding the 
set of source coefficients, $S_j$, that minimize the $\chi^2$.  Here, 
$\chi^2=\sum_i(C_i - \sum_j K_{ij} S_j)^2/\Delta^2C_i$.
This set of source coefficients is 
$S_j=\sum_i[(K^T(\Delta^2C)^{-1}K)^{-1}K^TB]_{ji} (C_i-1)$ 
where $K^T$ is the transpose of the kernel matrix.
The uncertainty of the source is the square-root of the diagonal elements of the
covariance matrix of the source,
$\Delta^2S=(K^T(\Delta^2C)^{-1}K)^{-1}$.
 \begin{figure}
\begin{center}
\vspace{-1cm}
 \epsfig{file=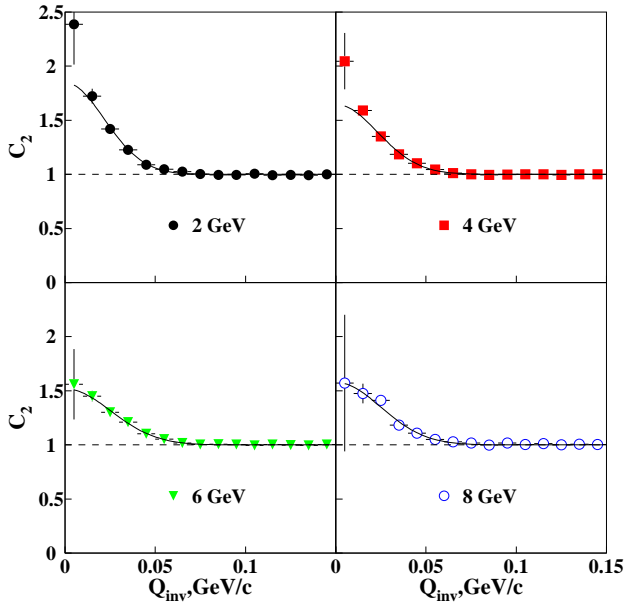,width=9.0cm}
 \caption{Measured two-pion correlation functions for Au + Au central
 collisions at four beam energies. Lines represent Gaussian fits to
the data.}
 \label{panitkin_1}
\end{center}
 \end{figure}
For the case of noninteracting spin-zero bosons with Gaussian correlations, 
there is a natural connection between the imaged sources and ``standard'' 
HBT parameters~\cite{dbrown_00_1} --- the source is a Gaussian with 
the ``standard'' HBT parameters as radii:
\begin{equation}
        S({\bf r})=\frac{\lambda}{(2\sqrt{\pi})^3\sqrt{\det{[R^2]}}}
        \exp{\left(-\frac{1}{4}r_ir_j[R^2]^{-1}_{ij}\right)}\,,
\label{s_gauss}
\end{equation}
Here $\lambda$ is a fit parameter traditionally called the chaoticity 
or coherence factor in HBT analyses and $[R^2]$ is the real symmetric 
matrix of radius parameters 
\begin{equation}
 [R^2]=\left(
\begin{array}{ccc}
                 R_o^2 & R_{os}^2 & R_{o\ell}^2 \\
                 R_{os}^2 & R_s^2 & R_{s\ell}^2 \\
                 R_{o\ell}^2 & R_{s\ell}^2 & R_\ell^2
\end{array}
\right).
\end{equation}
Eq.~(\ref{s_gauss}) is the most general Gaussian one may use, but 
usually one assumes a cylindrical symmetry of the single particle
source so that there is only one non-vanishing non-diagonal element 
$R_{ol}^2$.
The asymptotic value of the relative source function at zero 
separation $S({\bf r}\rightarrow 0)$ is related to the inverse 
effective volume of particle emission, and has units of fm$^{-3}$:
\begin{equation}
        S({\bf r}\rightarrow
        0)=\frac{\lambda}{(2\sqrt{\pi})^3\sqrt{\det{[R^2]}}} \,.
\label{connect}
\end{equation}

\begin{figure}
 \begin{center}
\vspace{-1cm}
 \epsfig{file=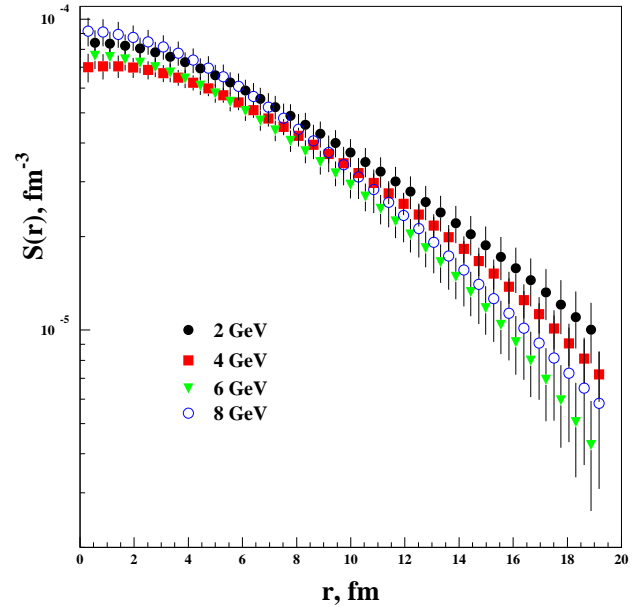,width=9.0cm}
 \caption{ Relative source functions extracted from the pion
 correlation data at 2, 4, 6 and 8$A$ GeV. }
 \label{panitkin_2}
 \end{center}
 \end{figure}

\begin{figure}
 \begin{center}
\vspace{-1cm}
 \epsfig{file=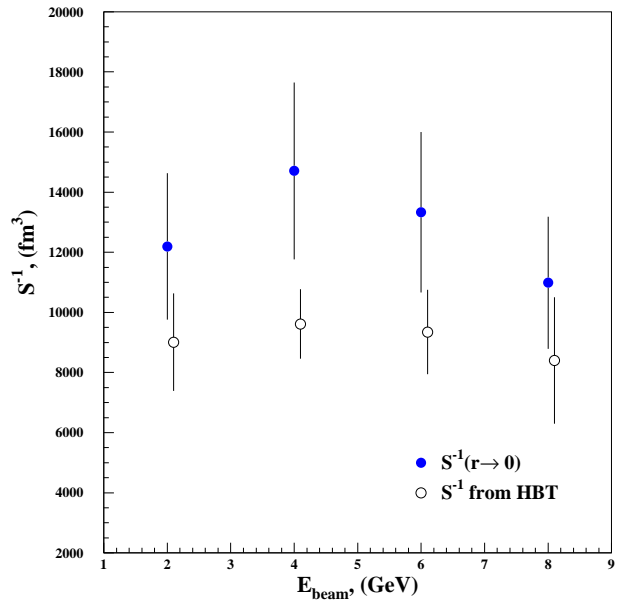,width=9.0cm}
 \caption{Values of inverse relative source functions at zero separation as a
 function of beam energy, obtained using imaging (solid circles) and HBT (open
circles).}
 \label{panitkin_3}
 \end{center}
 \end{figure}

As it was shown in~\cite{dbrown_00_1} $S({\bf r}\rightarrow 0)$ is an
important parameter needed to extract the space-averaged phase-space density. 
Figure~\ref{panitkin_1} shows measured angle-averaged 
two-pion correlation functions 
for central Au + Au collisions at 2, 4, 6 and 8$A$ GeV.
Figure~\ref{panitkin_2} shows relative source functions $S(r)$ 
obtained by applying the imaging technique to the measured two-pion 
correlation functions.  
Note that the plotted points are for representation of the continuous
source function and hence are not statistically 
independent of each other as the source functions are expanded 
in Basis Splines~\cite{dbrown_in_prep}. Since the source covariance
matrix is not diagonal, the coefficients of the Basis spline expansion are also
not independent which is taken into account during $\chi^2$
calculations.
The sources shown in Figure~\ref{panitkin_2} are consistent with zero,
within errors, in the region past $\approx20$ fm.
Thus with this technique, we may reconstruct the distribution of 
relative pion separations with useful accuracy out to $r \sim 20$ fm.
The images obtained at each of the four E895 beam energies are rather 
similar in shape, and upon fitting with a Gaussian function, values of 
$\chi^2$ per degree of freedom between 0.9 and 1.2 are obtained.  
Results of fits to the source function and correlation functions are
shown in Table~\ref{panitkin_t_1}. One can see that source radii
extracted via both techniques are similar, further
confirming the validity of a Gaussian source hypothesis.
Indeed, the somewhat smaller errors on the $R_S$ parameters compared to 
$R_{C2}$ are consistent with the Gaussian fit function being in closer 
overall agreement with the data points in the case of the imaging 
approach.
\begin{table}
\caption{Radius parameters of Gaussian fits to the extracted source
($R_S$) functions and measured correlation functions($R_{C2}$) for
different energies.}
\begin{tabular}{lcccc}
 $E_{b}$ (AGeV)  & 2 & 4 & 6 & 8 \\
 \hline
 $R_{S}$(fm)    &6.70$\pm$0.04& 6.35$\pm$0.03  & 5.56$\pm$0.03 & 5.53$\pm$0.05\\
 $R_{C2}$ (fm)  &6.39$\pm$0.2& 6.05$\pm$0.1  & 5.51$\pm$0.15  & 5.61$\pm$0.28 \\
\end{tabular}
\label{panitkin_t_1}
\end{table}  
 Figure~\ref{panitkin_3} compares the effective volumes of
pion emission inferred from the standard Bertsch-Pratt pion HBT 
parametrization (open circles) with the effective volumes derived from the 
image source functions shown in Fig.~\ref{panitkin_2} (solid circles),
recalling that $S({\bf r}\rightarrow 0)=S(r\rightarrow 0)$.  
Essentially, this figure plots the inverse of the left- and right-hand 
sides of Eq.~\ref{connect}. 
It can be seen from Fig.~\ref{panitkin_3} that the agreement between
imaging and the HBT parametrization is fairly good.  Values of source 
functions at zero separation as well as HBT source size parameters
(published in Ref.~\cite{lisa_1} and reproduced in Table~\ref{panitkin_t_2})
are approximately constant within errors across the 2 to 8$A$ GeV beam
energy range.
\begin{table}
\caption{Fit parameters for the Bertsch-Pratt HBT
parameterization of the pion correlation functions for E895 beam energies used for evaluation of $S(r\rightarrow 0)$.}
\begin{tabular}{lcccc}
 $E_{b}$ (AGeV)  & 2 & 4 & 6 & 8 \\
 \hline
 $\lambda$       &0.99$\pm$0.06& 0.74$\pm$0.03 & 0.65$\pm$0.03 & 0.65$\pm$0.05\\
 $R_o$  (fm)     &6.22$\pm$0.26& 5.79$\pm$0.16  & 5.76$\pm$0.23 &
5.49$\pm$0.31\\
 $R_{s}$ (fm)    &6.28$\pm$0.20& 5.37$\pm$0.11  & 5.05$\pm$0.12  & 4.83$\pm$0.21
\\
 $R_{l}$ (fm)    &5.15$\pm$0.19& 5.15$\pm$0.14  & 4.72$\pm$0.18 &
4.64$\pm$0.24\\
 $R_{ol}^2$ (fm)&-2.43$\pm$1.71& 0.43$\pm$1.03& 2.17$\pm$1.20 & -0.65$\pm$1.85
\\
\end{tabular}
\label{panitkin_t_2}
\end{table}
In summary, we present measurements of one-dimensional correlation functions 
for negative pions emitted at mid-rapidity from central Au + Au collisions 
at 2, 4, 6 and 8$A$ GeV.  These correlation functions are analyzed using the 
imaging technique of Brown and Danielewicz.  
It is found that relative source 
functions $S(r)$ have rather similar shapes, while zero-separation intercepts 
$S(r \rightarrow 0)$, which are related to the effective volume of pion 
emission, are almost constant across the range of bombarding energies under 
study.   Distributions of relative separation have been measured 
out to 20 fm, and the extracted source functions are approximately Gaussian.
We have performed the first experimental check of the predicted connection 
between imaging and traditional meson interferometry techniques and
found that the two methods are in good agreement.  Overall, results of this 
investigation strengthen the rationale for applying systematic imaging 
analyses to a variety of particle species (i.e., strongly interacting 
particles such as protons, antiprotons, etc., as well as pions), and open 
the door to more detailed inferences about spacetime structure.  Such 
inferences can play an important role in resolving outstanding questions 
about deconfined nuclear matter.  
Stimulating discussions with Drs.~G.F.~Bertsch, S.~Pratt, S.A.~Voloshin 
and N.~Xu are gratefully acknowledged.
This research is supported by US DOE, NSF and other funding, as detailed 
in Ref.~\cite{chung_00}.  
\nopagebreak

\end{document}